# ON THE QUANTUM MECHANICS FOUNDATION AND QUANTUM-CLASSICAL PHASE TRANSITION WITH SPONTANEPOS SUPERPOSITION BREAKING

Vladan Panković

Department of Physics, Faculty of Sciences, 21000 Novi Sad,
Trg Dositeja Obradovića 4. , Serbia , vladan.pankovic@df.uns.ac.rs

In this work we consider basic principles and problems of the standard quantum mechanical formalism. Especially we consider final measurement or detection procedure (collapse) as a quantum-classical continuous phase transition with spontaneous (non-dynamical) unitary symmetry (superposition) breaking (effective hiding). Practically, collapse as an effective (non-absolute) phenomena can be considered as an especial case of the general formalism of spontaneous symmetry breaking (with applications in many different domains of the physics, e.g. in elasticity of rigid bodies, quantum theory of ferromagnetism, quantum theory of electro-weak interactions as well as in chaotic inflation cosmology). General theory of the detection procedure (collapse) we consider in real experiment of single photon interference at beam splitter with final detection of the photon coordinate by usual photographic plate.

**Key words**: single photon interference, quantum-classical phase transition, spontaneous superposition breaking
**PACS number**: 03.65.Ta

## 1. Introduction. Basic principles and problems of the standard quantum mechanical formalism

As it is well-known quantum mechanics has been formulated between 1925. and 1927. practically one hundred years ago. It is well-known too that discussions about real physical meaning and problems of the standard quantum mechanical formalism are open to this day. In this work we shall consider basic principles and problems of the standard quantum mechanical formalism. Especially we shall consider final measurement or detection procedure (collapse) as a quantum-classical continuous phase transition with spontaneous (non-dynamical) unitary symmetry (superposition) breaking (effective hiding). Practically, collapse as an effective (non-absolute) phenomena can be consistently considered as an especial case of the general formalism of spontaneous symmetry breaking (with applications in many different domains of the physics, e.g. in elasticity of rigid bodies, quantum theory of ferromagnetism, quantum theory of electro-weak interactions as well as in chaotic inflation cosmology). General theory of the detection procedure (collapse) we shall consider in real experiment of single photon interference at beam splitter with final detection of the photon coordinate by usual photographic plate.

As it is well-known there are the following basic principles of the standard quantum mechanical formalism [1]- [6].

Quantum system QS is completely described or presented by corresponding quantum state of the unit norm from Hilbert space *HS* (QS).

Any physical characteristic of QS is completely described or presented by corresponding observable or Hermitian operator (with real eigen values) that acts in *HS* (QS).

Suppose that **A** represents some QS observable so that it is satisfied

(1) $\mathbf{A}| A_n > = A_n | A_n >$ for n=1,2, …

where $| A_n >$ represents eigen state and $A_n$ represents eigen value of **A** for n=1,2, … .

Then

(2) $B(\mathbf{A}) = \{ | A_n >, \text{for } n=1,2, \dots \}$

represents one referential frame in *HS* (QS). In mentioned referential frame quantum state $|\Psi >$ of QS has form

(3) $| \Psi > = \sum_n < A_n | \Psi > | A_n >$

where $< A_n | \Psi >$ for n=1,2, … represent superposition coefficients. It according to unit norm condition

(4) $< \Psi | \Psi > = 1$

implies

(5) $\sum_n |< A_n | \Psi >|^2 = 1$

Suppose that **B** represents some other QS observable with eigen states $| B_n >$ and eigen values $B_n$ for n=1,2, … and with corresponding referential frame $B(\mathbf{B}) = \{ | B_n >, \text{for } n=1,2, \dots \}$different from $B(\mathbf{A})$ (as it is case when **B** does not commute with **A**). In $B(\mathbf{B})$ quantum state $|\Psi >$ of QS has form

(6) $| \Psi > = \sum_n < B_n | \Psi > | B_n >$

under unit norm condition

(7) $\sum_n |< B_n | \Psi >|^2 = 1$

Then (3) as presentation of $| \Psi >$ in $B(\mathbf{A})$ has the same right as (6) as presentation of $| \Psi >$ in $B(\mathbf{B})$ for any **A** and **B** . In other words within quantum mechanics there is no any absolute referential frame.

As it is well-known within standard quantum mechanical formalism Schrödinger equation

(8) $\mathbf{H}| \Psi(t) > \equiv i\hbar d/dt \, | \Psi(t) >$

for initial quantum state

(9) $| \Psi(0) > = | \Psi_0>$

represents exact differential form of the exact quantum dynamical equation of QS. Here **H** represents Hamiltonian observable of QS, t – time moment, ħ- reduced Planck constant, and, $| \Psi(t) >$ – exact quantum mechanical dynamical state of given system in time moment t.

General solution of (8), (9), or integral form of exact quantum dynamics of QS, yields strictly deterministically quantum mechanical dynamical state in any time moment. It, formally, can be presented by

(10) $| \Psi(t) > \equiv \mathbf{U}(t) \, | \Psi_0> = \exp[\mathbf{H}t/(i\hbar)] \, | \Psi_0>$ .

where

(11) $\mathbf{U}(t) = \exp[\mathbf{H}t/(i\hbar)]$

represents unitary quantum mechanical dynamical evolution operator. In this way initial quantum state of QS strictly deterministically evolves during time.

In referential frame $B(\mathbf{A})$ it is satisfied

(12) $| \Psi(t) > \equiv \mathbf{U}(t) \, | \Psi_0> = \sum_n < A_n \, | \, \mathbf{U}(t) \, \Psi_0 > | \, A_n >$

which means that here superposition coefficients deterministically evolve during time without breaking of the general superposition form. Also in time dependent referential frame $B(\mathbf{U}(t)\mathbf{A}\,\mathbf{U}^{-1}(t))$ it is satisfied

(13) $| \Psi(t) > \equiv \mathbf{U}(t) \, | \Psi_0> = \sum_n < A_n \, | \, \Psi_0 > \mathbf{U}(t) | \, A_n >$

Here it is no any change of the initial superposition coefficients.

Simply speaking strictly deterministic unitary symmetric quantum dynamical evolution conserves superposition and does not admit existence of any absolute referential frame.

Suppose now that QS is described by quantum state (3). Suppose that measuring apparatus or detector D(**A**) for detection of the observable **A** acts at QS. Then, neglecting detailed analysis of the D(**A**) characteristics, there are effective experimental facts [7], [8] that QS immediately after detection is described by inherently statistical mixture of quantum states $| A_n >$ with corresponding statistical weights $|< A_n \, | \, \Psi >|^2$ for n=1,2, … . In other words during detection quantum superposition (3) becomes effectively broken, or it collapses and turns out in mentioned statistical mixture.

Obviously mentioned collapse cannot be explained by any quantum dynamical evolution of QS quantum states exclusively.

It is possible, however, consider quantum dynamical interaction between QS and quantum mechanically described D(**A**) [1].

Suppose that QS is initially, i.e. before quantum dynamical interaction with D**(A)**, described in *B***(A)** by the following quantum state

(14) $| \Psi > = \sum_n c_n | A_n >$

where $c_n$ for n=1,2, … represent superposition coefficients.

Suppose that **P(A)** represents so-called D**(A**) "pointer" observable that acts in Hilbert space *HS* (D(**A**)) of D(**A**) quantum states so that it is satisfied

(15) $\mathbf{P(A)} | P(A)_n > = P(A)_n | P(A)_n >$ for n=0,1,2, …

where $| P(A)_n >$ represent eigen states and $P(A)_n$ represent eigen values of **P(A)** for n=0,1,2, … .

Then

(16) $B(\mathbf{P(A)}) = \{ | P(A)_n >, \text{for } n=0,1,2, \dots \}$

represents one referential frame in *HS* (D(**A**)).

Suppose that D(**A**) is initially, i.e. before quantum dynamical interaction with QS, described in *B*(**P(A)**) by quantum state $| P(A)_0 >$.

In this way quantum super-system QS+D(**A**) (that includes both QS and D(**A**) ) is described by non-entangled quantum state $| \Psi >| P(A)_0 >$ from Hilbert space *HS* (QS+D(**A**))=*HS* (QS)**x***HS* (D(**A**)) of quantum states of mentioned super-system.

Suppose that quantum dynamical interaction between QS and D(**A**) is defined by unitary evolution operator **U(**QS+D(**A**)**)** that in general case satisfies condition

(17) $\mathbf{U}(QS+D(\mathbf{A})) | A_n >| P(A)_0 > = | A_n >| P(A)_n >$ for n=1,2, …. .

Then immediately after quantum dynamical interaction between QS and D(**A**) initial non-entangled quantum state of the super-system turns out strictly deterministically in the following entangled quantum state

(18) $| QS+D(\mathbf{A}) > = \mathbf{U}(QS+D(\mathbf{A})) | \Psi > | P(A)_0 > =$
$= \sum_n c_n | A_n >| P(A)_n >$

(whose existence can be metaphorically called Schrödinger cat effect) presented in referential frame

(19) $B(\mathbf{AxP(A)}) = \{ | A_n >| P(A)_m >, \text{for } n=1,2, \dots \text{ and } m=0,1,2, \dots \}$

Obviously entangled quantum state (18) represents an especial reduced form of the general form of the superposition

(20) $\sum_{n,m} d_{nm} | A_n >| P(A)_m >$ .

presented in referential frame *B*(**AxP(A)**) where $d_{nm}$ for n=1,2, …. and m=0,1,2, … represent superposition coefficients. In other words entangled quantum state (18) represents bi-orthogonal expansion of the quantum state of QS+D(**A**) over eigen basis of measured observable and correlated eigen basis of "pointer" observable.

Simply speaking quantum dynamical interaction between QS and D(**A**) does not break superposition of QS states (14) but it, without any change of the superposition coefficients, extends mentioned superposition at superposition (entanglement) of QS+D(**A**) states (18) .For this reason mentioned quantum dynamical interaction between QS and D(**A**) is not sufficient to describe effective experimental facts that QS immediately after detection is described by inherently statistical mixture of quantum states $| A_n >$ with corresponding statistical weights $|c_n|^2$ for n=1,2, … .

It can be supposed (von Neumann projection postulate [1]) that mentioned effective results of the final detection of **A** at QS simply correspond to inherently statistical mixture of quantum states $| A_n >| P(A)_n >$ with corresponding statistical weights $|c_n|^2$ for n=1,2, … . In other words it can be ad hoc postulated that after quantum dynamical interaction between QS and D(**A**) there is some different (non-unitary but projection) process according to which entangled quantum state (18) absolutely collapses or turns out in mentioned statistical mixture which prefers *B*(**AxP(A)**) as ultimate or absolute referential frame.

It can be hypothetically supposed within so-called hidden variables theories [9], [10] that absolute collapse can be considered as a dynamical breaking of the unitary symmetry of the quantum dynamics. Concretely such collapse is caused by dynamical terms (hidden variables) effectively hidden at less accurate quantum level of analysis but really existing at hypothetical more accurate level of analysis. Or, it can be supposed that hidden variables characterize QS, or D(**A**), or QS+D(**A**). But it is theoretically [11] and experimentally [12], [13] shown that in all mentioned cases hypothetical absolute collapse (as dynamical breaking of the unitary symmetry by hidden variables) implies necessarily extremely non-plausible super-luminal dynamical effects. On the other hand unitary symmetric quantum mechanical dynamics is always without any super-luminal effect. (This situation is similar to situation within quantum field theory where unification of the quantum electrodynamics and weak nuclear interaction cannot be realized by mass dynamical terms that break gauge symmetry of interactions and imply non-renormalizability [14]- [16].)

Within environmental or decoherence theories [17], [18] it has been supposed that effects of absolute collapse can be changed by dynamical or thermodynamical interaction between quantum system and its physical environment. Within decoherence theories there are the following statements. "Decoherence cannot therefore solve the phylosophical problem the understanding human perception of a particular reality. However it can explain the emergence of classicality, that is how and when an object loses its quantum features and becomes indistinguishable from a classical description.
The crucial point is to acknowledge that no quantum object is completely isolated, rather it is embedded in an environment consisting of gas particles, photons and the like. The environmental state gets easily "entangled" with the quantum object, which causes information about the whereabouts of the quantum object to be rapidly disseminated into surroundings. The absence of quantum behavior in the macroworld is a natural consequence of the fact that bigger and more complex objects are much harder to isolate. In other words, the quantum features of the environmental interactions and the resulting information transfer lead to the appearance of classicality in quantum objects.” [7]

But as it is well-known and as it has been discussed previously, according to standard quantum mechanical formalism [4] and experimental facts the following is satisfied. Firstly, entangled quantum state of the quantum super-system with any number of the quantum sub-systems is strictly different from statistical mixture of the non-entangled quantum states of quantum sub-systems. Secondly, only after so-called sub-systemic detection, when detector (e.g. photographic plate) detects only one quantum system (e.g. coordinate of large molecule after transition through a diaphragm with two slits) entangled with environment (e.g. gas of many small molecules between diaphragm and photographic plate that later do not interact with photographic plate) quantum system is described by sub-systemic statistical mixture (of the first kind). After mentioned entanglement with environment but before sub-systemic detection mentioned quantum system is not really physically described by statistical mixture of the first kind (even if formally, but not really, it can be described by so-called statistical mixture of the second kind with mathematical form identical to statistical mixture of the first kind). So, decoherence, as an absolute breaking of the superposition of quantum states of quantum system already and only by entanglement with environment, needs implicitly but necessarily introduction of some (thermo)dynamical hidden variables dependent of the temperature and number of the physical systems in environment. But of course such hidden variables must imply necessarily super-luminal dynamical effects. Finally there are new extremely important experimental facts [19] that point out that quantum entanglement stands conserved even for extremely large number of the entangled quantum systems and very large temperatures. is independent of the number of entangled systems and temperature of the environment .

Within so-called orthodox interpretation (within which there is no any classical mechanical concept) [1] von Neumann hypothetically supposed that since absolute collapse cannot be consequence of any physical dynamical evolution (von Neumann proved a no-go theorem for hidden variables, but this theorem has limited domain of validity), generator of the absolute collapse is absolutely meta-physical Abstract Ego of human observer. It in fact represents meta-physical hidden variable and as well as any other hidden variable that hypothetically causes absolute collapse it must acts necessarily superluminally.

So, concept of the absolute collapse within standard quantum mechanical formalism is deeply analogous to concept of the absolute space and time within classical mechanics [20], [21]. But it does not mean that detection results can be explained without effective collapse concept.

Namely bi-orthogonal expansion of | QS+D(**A**) > (18) is not necessarily unique so that, at least in some relevant cases [22], there is different bi-orthogonal expansion

$$(21) \quad | QS+D(\mathbf{A}) > = \sum_n g_n | B_n >| P(\mathbf{B})_n >$$

where $g_n$ for n=1,2, … represent superposition coefficients while **B** represents some observable of QS that does not commute with **A** and **P(B)** represents some observable of D(**A**) that does not commute with **P(A)**. In other words entangled quantum state (21) represents bi-orthogonal expansion of the quantum state of QS+D(**A**) over eigen basis of measured observable **B** and correlated eigen basis of “pointer” observable **P(B)**. It is in full agreement with unitary symmetry of the quantum mechanical dynamics. In other words the same, unique quantum dynamical state of QS+D(**A**) admits many correct different forms of entanglement and only effective collapse prefers (retrospectively) one correct form of entanglement. In such (retrospectively) preferred entanglement form it can be additionally hypothetically supposed elimination of the effective collapse by so-called sub-systemic splitting of the super-system (Universe branching) over relative, i.e. correlated quantum states [23]. It formally means change of the inherently statistical effective collapse with conservation of the unit norm of quantum states by deterministic splitting or "fission" of quantum states without conservation of the unit norm. All this contradicts to unitary symmetry of the quantum mechanical dynamics and experimental facts.

Also within standard quantum mechanical formalism, i.e. exactly quantum mechanically, in full agreement with experimental facts, superposition stands conserved even in case of weakly interfering quantum states of QS or D(**A**)**.** For this reason suppositions that collapse can be simply eliminated by weak interference (including relativistically distant) QS or D(**A**) quantum states [24], [25] are inconsistent. As it is well-known [1] in distinction to reversible unitary symmetric quantum dynamical evolution effective collapse is effectively irreversible. In [5] is, in fact, excellently demonstrated how by different positions of many magnets in Stern-Gerlach spin measurement spin and trajectory of a quantum system can be (reversibley) entangled and de-entangled before (irreversible) final detection by photographic plate.

Finally there is option, intuitively presented in [20], [21], and later detailedly formalized in [26]- [28], that effective collapse (without any absolute characteristics) represents a continuous quantum-classical phase transition with spontaneous (non-dynamical) unitary symmetry (superposition) breaking (effective hiding).

Generally speaking Bohr firstly intuitively supposed (without any consequent formalization) that at the exact quantum mechanical level of the analysis accuracy unitary symmetric quantum dynamics completely describes all physical phenomena. Also, in full epistemological analogy with theory of relativity [21], unitary symmetry of quantum mechanical dynamics means that all referential frames (eigen states bases of all observables) in Hilbert space have the same right so that there is no any absolute referential frame in the quantum mechanics (vice versa absolute observer in the quantum mechanics is conceptually analogous to absolute space and time in Newtonian classical mechanics). In this sense Bohr intuitively supposed that by quantum dynamical interaction between quantum system and detector (we paraphrase) "a non-separable unit is realized" which later has been consequently formalized by von Neumann entanglement between quantum system and detector.

Secondly Bohr intuitively, but without any consequent formalization, supposed that by detection the detector is described effectively classically. Only relatively, i.e. in respect to so-described detector quantum system is described effectively by statistical mixture of quantum states. All this appears effectively only at some approximate level of analysis accuracy, but without any breaking of the unitary symmetry at the exact quantum level of analysis accuracy. Strictly exactly the detection, i.e. superposition breaking, does not appear at all instead of which entanglement between quantum system and detector exactly exists. In this way there are two discretely different levels of the analysis accuracy, exact (quantum) and effective (classical), and either first or second can be chosen for description of the interaction between quantum system and detector. It probably represents best presentation of the Bohr concept of the movable boundary between quantum system and detector.(It can be metaphorically presented by Bohr words "that we are actors and spectators in the great drama of existence" (paraphrase).)

But how it can be consequently formalized , that is the question. Answer is simple, by general formalism of the spontaneous (non-dynamical) symmetry breaking (effective hiding). Practically, collapse as an effective (non-absolute) phenomena can be considered as an especial case of the general formalism of spontaneous symmetry breaking (with applications in many different domains of the physics, e.g. in elasticity of rigid bodies, quantum theory of ferromagnetism, quantum theory of electro-weak interactions as well as in chaotic inflation cosmology [14]- [16], [29]). Mentioned option on the collapse as the spontaneous superposition breaking by quantum-classical phase transition will be considered in next section of this work.

## 2. Quantum-classical phase transition with spontaneous superposition breaking

Consider, now, for some quantum system, quantum mechanical dynamical state from orbital Hilbert space. Orbital Hilbert space represents an especial sub-space of the usual Hilbert space. Over this sub-space only such observables act that represent analytical functions of the coordinate observable **x=**x or/and momentum observable **p**.

Then, as it is well-known [6] , Schrödinger exact quantum mechanical dynamical equation for quantum state can be changed by equivalent Ehrenfest exact quantum mechanical dynamical equations for average values

(22) $m\, d\langle \mathbf{x} \rangle/dt = \langle \mathbf{p} \rangle$

(23) $d\langle \mathbf{p} \rangle/dt = - \langle \partial \mathbf{V}(x)/\partial x \rangle$

Here m represents mass and **V**(x) - potential energy observable of quantum system.

Right hand of (23) can be Taylor expanded in a relatively small vicinity $\Delta x$ of $\langle x \rangle$, where $\Delta x$ represents standard deviation of x. It yields

(24)
$$\langle \partial \mathbf{V}(x)/\partial x \rangle = \partial \mathbf{V}(\langle x \rangle)/\partial \langle x \rangle + [\partial^2 \mathbf{V}(\langle x \rangle)/ \partial \langle x \rangle^2]\langle x-\langle x \rangle \rangle + (1/2)[\partial^3 \mathbf{V}(\langle x \rangle)/ \partial \langle x \rangle^3]\langle x-\langle x \rangle \rangle^2 + \dots$$
$$= \partial \mathbf{V}(\langle x \rangle)/\partial \langle x \rangle + (1/2)[\partial^3 \mathbf{V}(\langle x \rangle)/ \partial \langle x \rangle^3]\Delta x^2 + \dots .$$

In general case series (24) diverges. But series (24) can converge and first term in Taylor expansion becomes predominant for some well-known [6] quantum mechanical dynamical states called wave packets that belong to orbital Hilbert space.

Within wave-packet approximation standard deviation of any observable that acts over orbital Hilbert space (including coordinate and momentum observable) can be formally neglected in respect to absolute value of average value of mentioned observable. However as it is well-known [6] there is down limit of applicability of such approximation (including Heisenberg uncertainty relations).

Within satisfied wave packet approximation equation (22) stands practically unchanged while equation (23) turns out in the following equation

(25) $d\langle \mathbf{p} \rangle/dt = - \partial \mathbf{V}(\langle x \rangle)/\partial \langle x \rangle$

It corresponds, obviously, to Newton classical mechanical dynamics for a particle with coordinate $<x>$ and momentum $<\mathbf{p}>$ for classical mechanical force $-\partial \mathbf{V}(<x>)/\partial <x>$.

Exact quantum mechanical dynamical state $| \Psi >$ in orbital Hilbert space always exists. But only when this quantum state satisfies all mentioned approximation conditions it represents approximately a wave packet $| \Psi, WP >$ placed in small vicinity of $<x>$ as it is symbolically presented at figure 1.a).

It is very important that the following to be observed and pointed out. During continuous unitary dynamical evolution of always existing exact quantum mechanical dynamical state $| \Psi >$ during time, cases of satisfaction of wave packet approximation conditions and wave packet $| \Psi, WP >$ existence can continuously turn out in cases of non-satisfaction of wave packet approximation conditions and wave packet $| \Psi, WP >$ non-existence (e.g. by wave packet dissipation) and vice versa. All this can be formally considered as corresponding continuous phase transitions.

Thus, exact quantum mechanical dynamics is always exactly globally stable in the orbital Hilbert's space. Also, when wave packet approximation is satisfied, quantum mechanical dynamics can be presented effectively by classical mechanical dynamics globally stable within usual coordinate space, i.e. phase space. Vice versa, when wave packet approximation is unsatisfied then exact quantum mechanical dynamics cannot be approximated classical mechanically dynamically globally stablely in the usual coordinate space, i.e. phase space. Finally, transition from the situation when wave packet approximation is satisfied in situation when wave packet approximation is unsatisfied (and vice versa) can be presented as a typical continuous phase transition.

Consider, now, for some quantum system, two quantum states $| \Psi_1 >$ and $| \Psi_2 >$ from orbital Hilbert space so that quantum dynamical state of mentioned quantum system represents some superposition of mentioned two quantum states. As it has been previously discussed mentioned exactly existing superposition stands always conserved by mentioned quantum dynamical evolution during time.

Suppose that both quantum states $| \Psi_1 >$ and $| \Psi_2 >$ satisfy wave packet approximation so that wave packets $| \Psi_1, WP>$ and $| \Psi_2, WP>$ with corresponding coordinate average values and standard deviations $<x_1>$ and $\Delta x_1 = \Delta x$ and $<x_2>$ and $\Delta x_2 = \Delta x_1 = \Delta x$ simultaneously exist. However, it can be simply proved that when mentioned two wave packets weakly interfere then superposition of mentioned wave packets does not represent any wave packet. Consider for simplicity superposition when absolute vale of superposition coefficients are proportional or practically almost equivalent. Namely then absolute value of the coordinate average value in superposition of wave packets equals practically $|(<x_1> + <x_2>)/2|$ while coordinate standard deviation in superposition of wave packets equals practically $|(<x_1> - <x_2>)/2|$. Further we can consider $|(<x_1> - <x_2>)|$ as order parameter with critical value $\Delta x$ so that for order parameter smaller than critical value two wave packets overlap and interfere strongly, while in opposite case, i.e. for order parameter larger than critical value two wave packets weakly interfere as it is symbolically presented at figure 1.b) and 1.c). In last case of weakly interfering two wave packets superposition of mentioned wave packets does not satisfy all wave packet conditions (distribution of the quantum system coordinate nearly $|(<x_1> + <x_2>)/2|$ can be neglected, Taylor expansion of Ehrenfest average value of the quantum dynamical equation becomes divergent, etc.). For this reason exactly existing quantum superposition of two or more weakly interfering wave packets,

approximately classically is globally non-stable and it cannot be globally presented as any wave packet.

However, mentioned superposition effectively can be locally classically presented either by one (“left”) or by other (“right”) wave packet in corresponding small coordinate domains in which mentioned wave packets are locally classically well-defined. This situation is conceptually completely analogous to situation by spontaneous gauge symmetry breaking in electro-weak interactions [14]- [16].

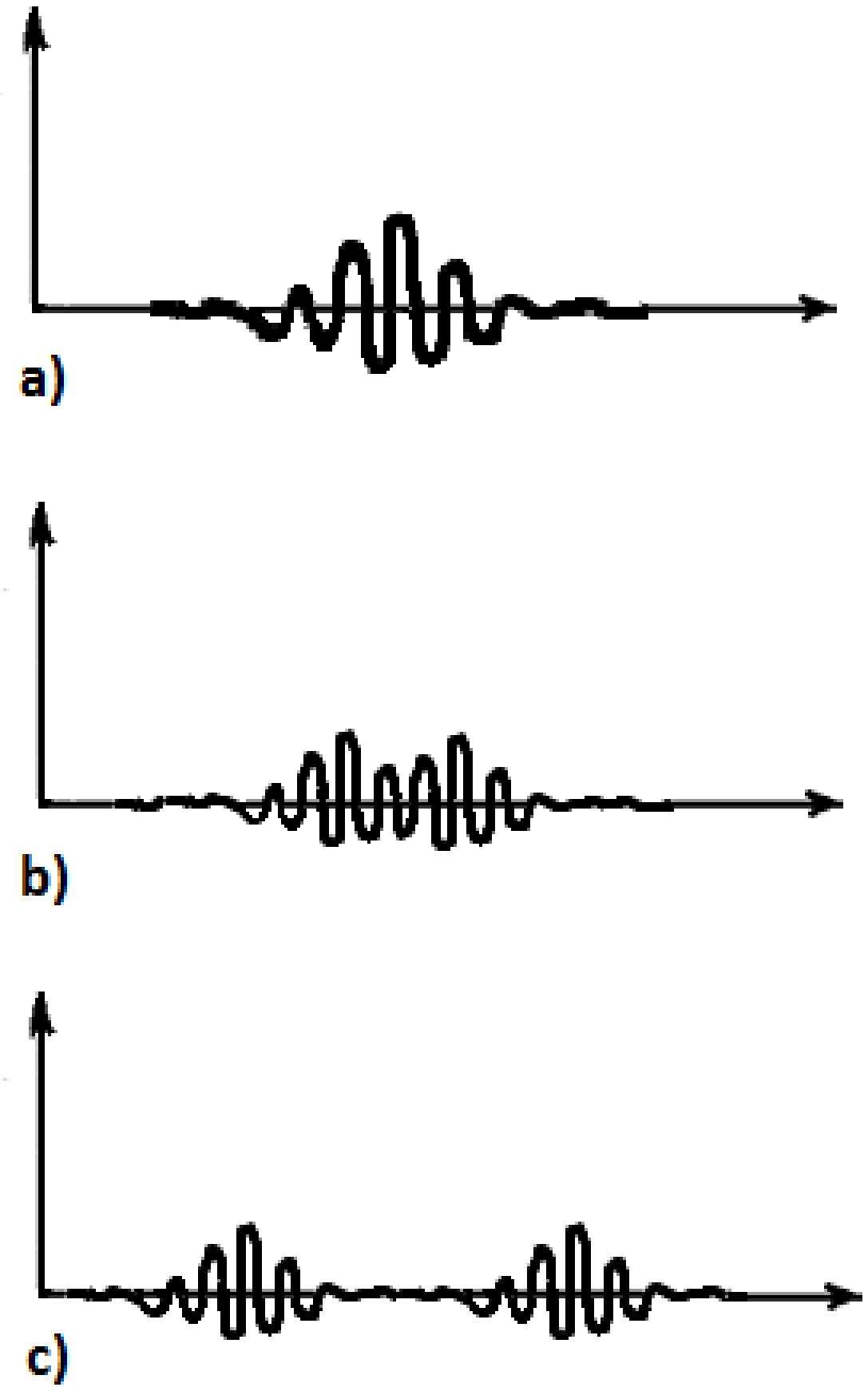

**Figure 1.**

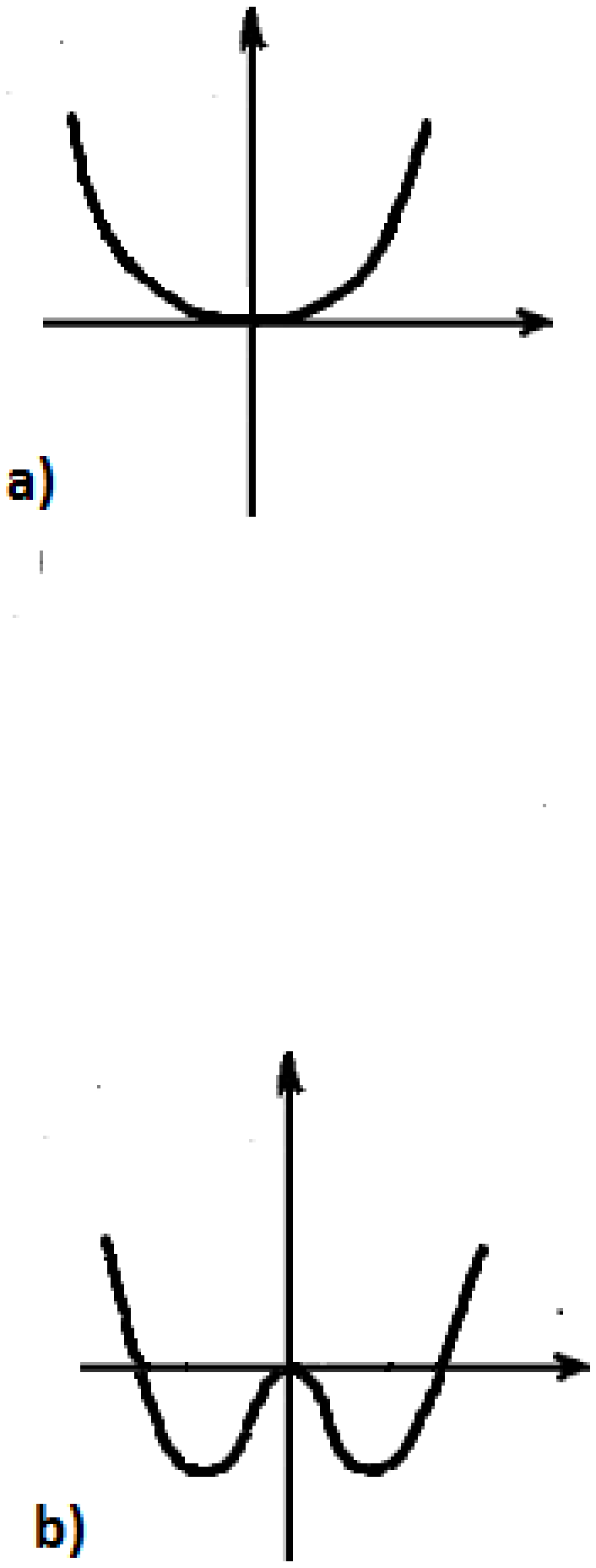

**Figure 2.**

As it is well-known theory of electro-weak interaction must be exactly gauge symmetric since only in this case mentioned theory admits renormalization. For this reason boson masses cannot be obtained by introduction of additional dynamical terms which exactly break gauge symmetry. However boson masses can be obtained by spontaneous (non-dynamical) breaking (effective hiding) of the gauge symmetry. Simply speaking, exact dynamical equations of the quantum field theory for electro-weak interactions must be solved iteratively approximately, by perturbation theory (which has character of a Taylor expansion).Within false vacuum quantum field point perturbation theory globally diverges, while within a circular ring of real vacuum quantum field points perturbation theory locally converges. For this reason from perturbation theory approximate level of accuracy, dynamical characteristics of the false vacuum are dynamically globally non-stable and non-describable, while dynamical characteristics of the real vacuums are dynamically locally stable and describable. Formally and effectively false vacuum can be changed by one of many equally probable real vacuums when circular symmetry of the real vacuums as well as gauge symmetry becomes spontaneously (non-dynamically) broken (effectively hidden). Mentioned change and irreduciblely statistical choice characterize in fact perturbation theory and have not absolute character (mentioned choice of some real vacuum is not governed by some small additional dynamical term). ( All this is, in usual analogy [16], symbolically presented at figure 2. for scalar fields. At figure 2.a) it is presented potential energy density for scalar field with positive mass term and single global real vacuum. At figure 2.b) it is presented potential energy density for self-interacting $\varphi^4$ scalar field with negative mass term with single false vacuum and two, “left” and “right”, local real vacuums. Here, from perturbation theory approximate level of analysis accuracy, “left”-“right” symmetric false vacuum effectively turns out inherently statistically either in “left” or in “right” local real vacuum with spontaneous breaking of “left”-“right” symmetry.)

All this admits formulation of the general formalism of the spontaneous (non-dynamical) symmetry breaking (effective hiding) in the following way. Simply speaking by dynamical symmetry breaking in some of many different physical theories dynamical symmetry at some level of analysis accuracy is broken by dynamical terms effectively hidden at this accuracy level but really existing at more exact accuracy level. It can be added that dynamical symmetry breaking represents physical content for many usual gambling games, e.g. at play at dice etc. . It can imply a seeming impression that dynamical symmetry breaking represents unique physical corner stone for mathematical theory of probability. By spontaneous symmetry breaking in some of many different physical theories (e.g. in elasticity of rigid bodies, quantum theory of ferromagnetism, quantum theory of electro-weak interactions as well as in chaotic inflation cosmology [14]- [16], [29]) dynamical symmetry and dynamical state really exist at exact level of analysis accuracy. However, dynamical state can be expressed only approximately, by some iterative, e.g. Taylor expansion series including perturbation theories. Mentioned Taylor series can converge not in all domains, i.e. generally or globally but only in some different domains, i.e. particularly or locally. For this reason exactly existing symmetric dynamical state becomes effectively hidden or effectively dynamically non-stable for approximate iterative level of analysis accuracy. It can be locally projected or changed by some particular dynamical state that is locally effectively dynamically stable within corresponding local domain of Taylor series convergence. Simultaneously mentioned change implies effective hiding of the symmetry. Obviously mentioned projection of effectively dynamically non-stable state in effectively dynamically stable state is effectively irreversible and it cannot be completely described neither by exact nor by effective dynamics. However, mentioned projection can be effectively considered as inherently statistical or probabilistic event. Of course, after realization of mentioned event effective dynamical state evolves strictly deterministically according to effective dynamics. All this is metaphorically and epistemologically analogous to well-known story about competition between three taylors in the same street in some town. First taylor stated that he is best taylor in the town. Second taylor stated that he is the best taylor in the world. But the third taylor, winner, stated only that he is the best taylor in this street.

According to general formalism of the spontaneous symmetry breaking it can be supposed [27], [28]) that superposition of weakly interfering wave packets, within classical mechanical approximation only, effectively self-collapses or turns out spontaneously (non-dynamically) in statistical mixture of mentioned wave packets. On the other hand, at exact quantum level of analysis accuracy, there is no any breaking of the superposition of corresponding wave packets. Self-collapsed state, at approximate level of analysis accuracy, evolves further classical mechanical dynamically deterministically and it cannot here return in initial approximately classical dynamically globally non-stable state. On the other hand at exact quantum level of analysis accuracy where entangled quantum state exists self-collapse has been never realized and exact quantum phenomena are reversible. It simply explain previously discussed principal difference between reversible exact unitary symmetric quantum dynamical evolution and effectively classically irreversible collapse [1]. All this admits metaphoricall statement that (self)collapse represents cut of previously mentioned third taylor.

Since any superposition of weakly interfering wave packets effectively self-collapses it implies that basis of all weakly interfering wave packets has within mentioned approximation especial, effectively dominant role so that it can be formally considered as “absolute” referential frame and self-collapse as “absolute” collapse. In this way weakly interfering wave packets can be effectively considered as the eigen states of the “absolute” “pointer” observable.

Self-collapse as continuous phase transition with spontaneous superposition breaking can be presented in the following way. Suppose that initially all wave packets strongly interfere (as it is symbolically presented at figure 1.a) ). Suppose that, during quantum dynamical evolution, distance between any two wave packet centers (average coordinate values) as order parameters increase while coordinate deviation of any wave packet stands practically the same (wave packets non-dissipation condition). Then during time any two wave packets interfere weaker and weaker (as it is symbolically presented at figure 1.b) ) and finally all wave packets become mutually weakly interfering or practically non-interfering (as it is symbolically presented at figure 1.c) ) when self-collapse appears. All this additionally is symbolically presented at figure 3. (analohous with figure 1.) for superposition of two wave packets with the same superposition coefficients. (where QM EX denotes quantum mechaucally exact description and CM APP – classical mechanical approximate description, while 1/2 denotes stastistical weight).

So, self-collapse as the spontaneous superposition breaking can be realized only at quantum states from orbital Hilbert space that represent superposition of weakly interfering wave packets during time intervals within which all wave packet approximation conditions are satisfied. It means that self-collapse can be in principle realized at any physical system, microscopic or macroscopic. However for microscopic systems with relatively small masses wave packets dissipate very quickly while for macroscopic systems with relatively large masses wave packets dissipate very slowly. For this reason classical mechanical description of microscopic system or quantum mechanical description of macroscopic systems are technically extremely hard for realization even if both, first effectively and second exactly, are possible in principle.

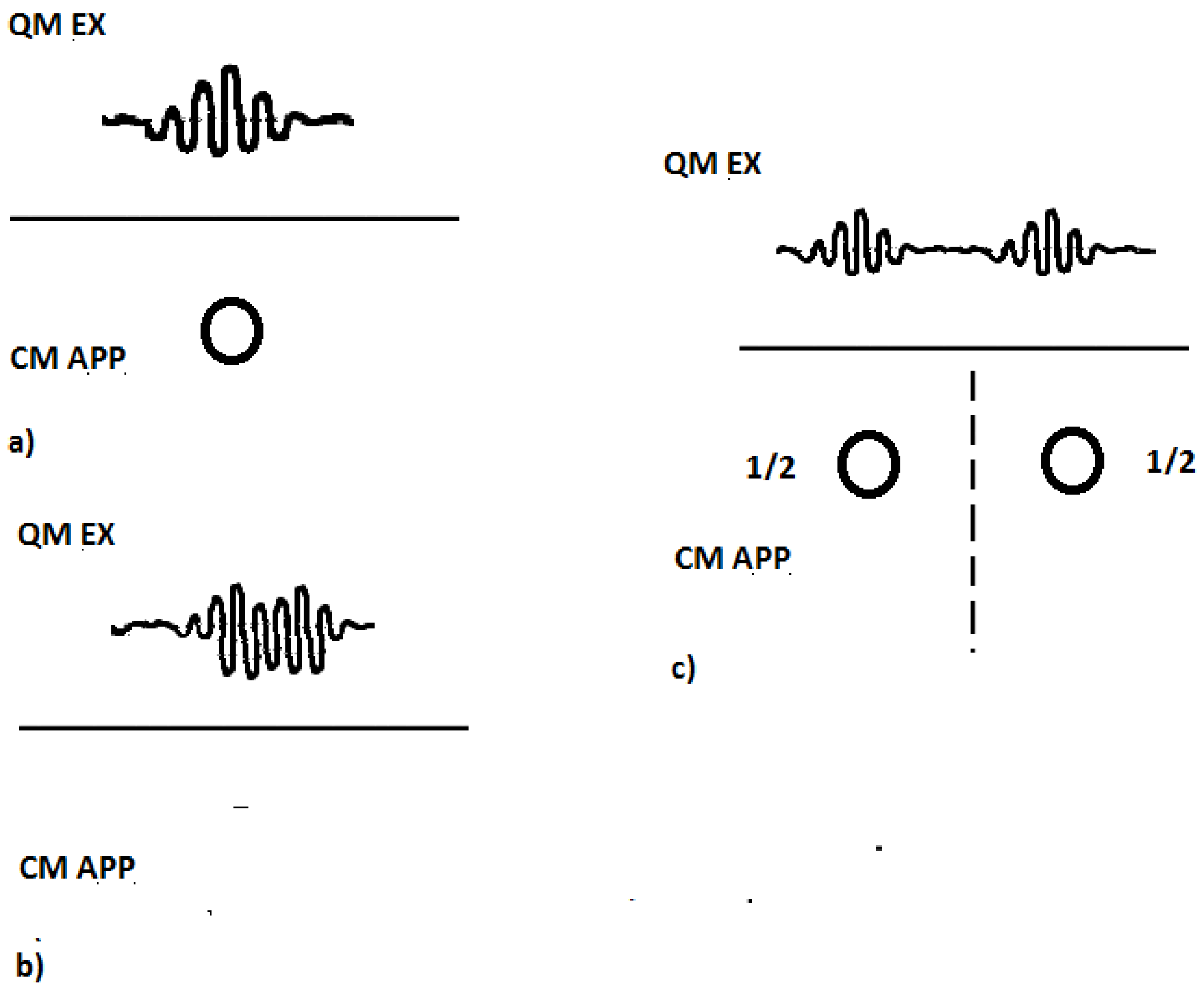

Figure 3.

Quantum states that do not belong to orbital Hilbert space of some quantum system , “object”, cannot realize self-collapse at all. Also, when conditions for self-collapse are not satisfied, “object” quantum state from orbital Hilbert space cannot realize self-collapse. However, “object” can quantum dynamically interact with appropriate other quantum system, “detector” at which self-collapse can effectively occur. Suppose that by mentioned quantum dynamical interaction the entanglement between “detector” and “object” is exactly realized. It includes realization of the exactly existing entangled quantum state bi-orthogonally expanded over wave packet approximated eigen basis of “detector” “pointer” observable and eigen basis if some “object” observable. In this case when at “detector” self-collapse effectively appears, i.e. when some eigen state of the “detector” pointer observable appears with corresponding statistical weight, then, via entanglement, at “object” correlated eigen state of mentioned observable appears with (conditional) certainty. In this sense we can speak about relative collapse at “object” as effectively pure quantum phenomena.

It represents general theory of the detection procedure (collapse) as a quantum-classical continuous phase transition with spontaneous (non-dynamical) unitary symmetry (superposition) breaking (effective hiding).

Presented general theory of the detection procedure (collapse) as a quantum-classical continuous phase transition with spontaneous (non-dynamical) unitary symmetry (superposition) breaking (effective hiding) we shall consider in real experiment of single photon interference at beam splitter with final detection of the photon coordinate by usual photographic plate in next section of this work.

### 3. Single photon interference at beam splitter. Conclusion

As it is well-known remarkable experiments of the interference of single quantum system (photon, electron, atom, molecule, etc.) at a diaphragm with two slits or, analogously, at beam splitter, etc. [2],[5], [20], [30]-[32] (in excellent agreement with contemporary sophisticated real experimental facts [7], [8]) represent corner stone for demonstration of the basic principles and problems (collapse or superposition breaking) of standard quantum mechanical formalism.

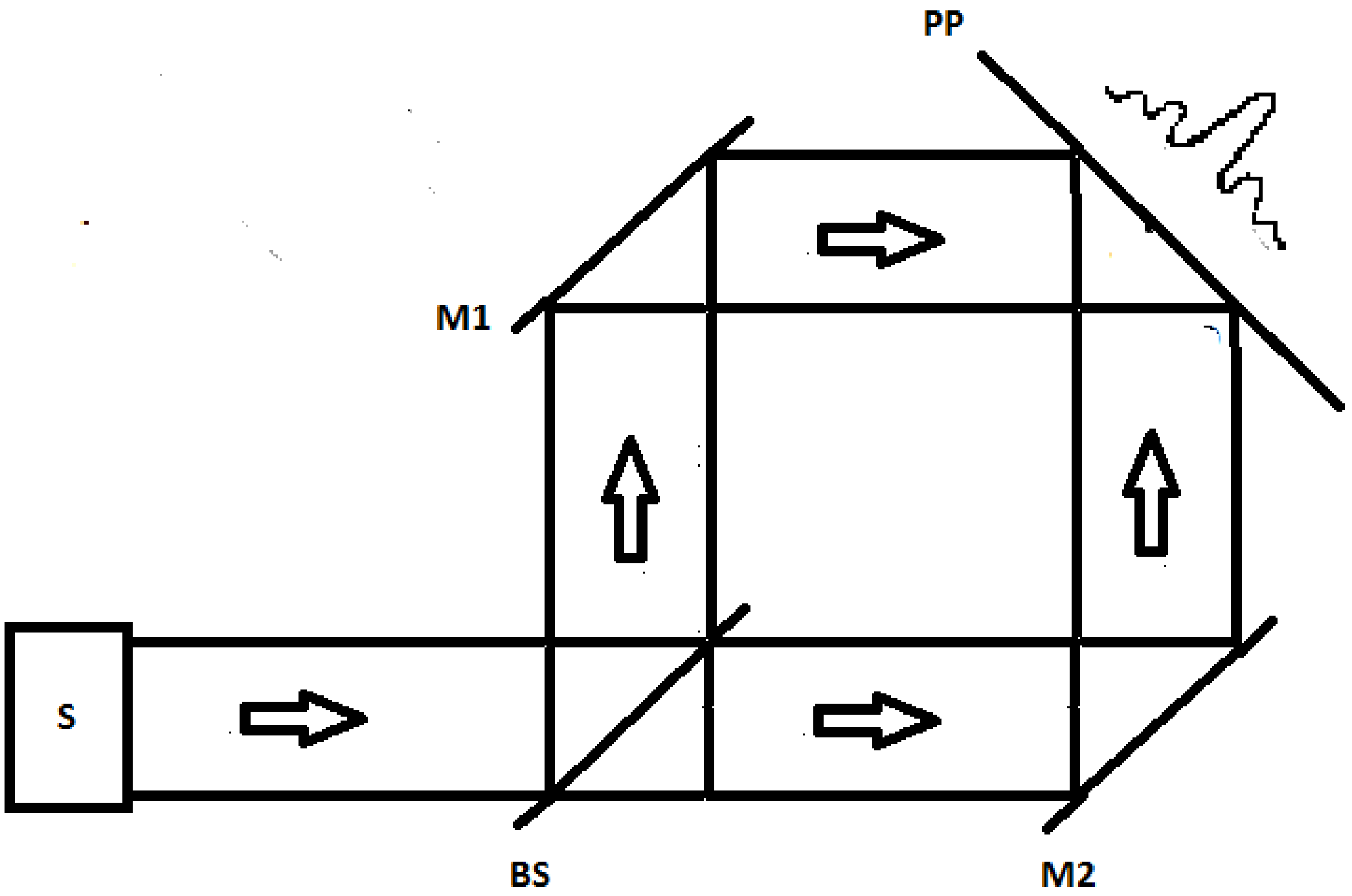

**Figure 4.**

Consider usual experiment of single photon interference at beam splitter without photon trajectories detection symbolically presented at figure 4. . Here source S emits single photon (wavefront) with well-defined momentum and corresponding wave length that propagates toward beam splitter BS. "Reflected" photon (redirected by total mirror M1) propagates toward photographic plate P**P** that represents photon coordinate detector, while "transmitted" photon (redirected by total mirror M2) propagates toward PP too.

Then, immediately before detection that PP realizes at photon, photon is described by the following quantum superposition state

(26) $|ph > = c_1 |R > + c_2 |T >$

Here $|R >$ and $|T >$ represent "reflected" and "transmitted" quantum state (trajectory) while $c_1$ and $c_2$ represent superposition coefficients (which includes especial case characteristic for usual experiment where mentioned superposition coefficients equal $i\ 2^{-1/2}$ and $2^{-1/2}$ respectively). Final detection of single photon (coordinate), i.e. absorption of single photon can occur in any "point", i.e. elementary segment (part) of PP.

We shall now accurately consider how PP quantum dynamically interacts with single photon described by (26).

Suppose that PP (without any diminishing of the basic conclusions generality) can be presented by the following simplified model.

Suppose that PP represents one-dimensional physical system with characteristic space direction q. Also suppose that PP consists of many identical, quantum dynamically practically non-interacting elementary segments EPP1, EPP2, … , EPPn placed in series along q direction. Suppose that in any elementary segment of PP there is a relatively large number of the photo-sensitive micro-elements, e.g. photo-sensitive micro-crystals of the same type of silver halide which are practically homogeneously distributed. Finally suppose that EPPn has a small length $\Delta q$ and that EPPn occupies coordinate interval
$I_n(q) = ((n-1)\Delta q , n\Delta q )$ with centers in $q_n = (n-1/2)\Delta q$ for n=1,2, … .

Define $|n,0 >$ as quantum state of EPPn without absorption of any photon for n =1,2, …..

Define $|n,1 >$ as quantum state of EPPn after absorption of single photon for n =1,2, ….. . By absorption of single photon photo-sensitive micro-element changes principally its characteristics. For example, as it is well-known, absorption of single photon by silver halide micro-crystal corresponds to transition of corresponding single electron from valence band (where mentioned electron is localized nearly some atom in crystal lattice) in conduction band (where mentioned electron is delocalized and can move through crystal). Such delocalized electron is attracted toward an electron trap (crystalline defect etc.) at sensitivity speck at crystal surface. Here finally electron is recombined with some interstitial silver ion in neutral silver atom (in common with corresponding energy, i.e. photon emission). It can be observed that initial quantum state of electron localized nearly some atom in crystal valence band and final quantum state of electron localized in silver atom in sensitivity speck extremely weakly interfere. For this reason $|n,0 >$ and $|n,1 >$ extremely weakly interfere too for n=1,2, … .

Define "zero" eigen state of PP "pointer" observable

(27) $|PP,0 > = |1,0 >|2,0 > \dots | n,0 >$ .

Define n-th eigen state of PP "pointer" observable

(28) $|PP,n > = |1,0 >|2,0 > \dots|n-1,0 >| n,1 >| n+1,0 > \dots$
for n =1,2, …..

Finally suppose that PP before quantum dynamical interaction with some photon ph is described by quantum state (27) while simultaneously ph is described by some quantum state $|ph,p,X >$ (which also includes superposition of different photon trajectories). It implies that quantum super-system ph+PP, before quantum dynamical interaction between ph and PP, is described by non-entangled quantum state
$|ph,p,X >|PP,0 >$.

Define $|ph,q,n >$ as the quantum state of ph when q coordinate of ph belongs to coordinate interval $I_n(q)$ for n=1,2, … .

Then quantum state of ph, immediately before quantum dynamical interaction between ph and PP, can be presented by the following expression

(29) $$|ph,p,X > = \sum_n < ph,q,n |ph,p,X > |ph,q,n >$$

that generally represents superposition of quantum states $|ph,q,n >$ with superposition coefficients $< ph,q,n |ph,p,X >$ for n =1,2, …. .

Define unitary symmetric evolution operator **U** corresponding to quantum dynamical interaction between ph and PP by expression

(30) $$\mathbf{U}|ph,q,n > |PP,0 > = |ph,q,n > |PP,n >$$

for n =1,2, …..

Then, immediately after quantum dynamical interaction between photon and PP non-entangled quantum state $|ph >|PP,0 >$ turns out strictly deterministically in the following entangled quantum state

(31) $$|ph+PP > = \mathbf{U}|ph > |PP,0 > = \sum_n < ph,q,n |ph,p,X > |ph,q,n >|PP,n >$$

It can be observed that, in EPPn, initially captured electron in valence band, i.e. nearly some atom can be approximately described by a wave packet (initial wave packet) $|n,0,WP >$ without momentum average value, with coordinate average value corresponding to center of the domain of initial localization and wave-packet width corresponding to linear dimension of the domain of initial localization for n=1,2, … . Also, in EPPn, finally captured electron in neutral silver atom can be approximately described by a wave packet (final wave packet) $|n,1,WP >$ without momentum average value, with coordinate average value corresponding to center of the domain of final localization and wave packet width corresponding to linear dimension of the domain of final localization for n=1,2, … . Obviously, in EPPn, mentioned electron initial and final wave packets extremely weakly interfere for n=1,2, … . Then, in simple analogy , $|PP,n,WP >$ as wave packet approximation of $|PP,n >$ can be defined for n=0,1,2, … .Obviously wave packet approximations of all eigen states of PP "pointer" observable mutually extremely weakly interfere.

In mentioned partial wave-packet approximation entangled quantum state (31) turns out in the following entangled quantum state in partial wave-packet approximation

(32) $$\sum_n < ph,q,n |ph > |ph,q,n >|PP,n,WP >$$

According to previous discussions entangled quantum state (32) spontaneously collapses or turns out effectively in inherently statistical mixture of the states $|ph,q,n >|PP,n,WP >$ with probability or statistical weights

(33) $$|< ph,q,n |ph,p,X > |^2$$

for n=1,2, ….

Consider especially case when |ph,p,X > represents superposition of different quantum trajectories (26). Then (33) turns out in

$$(34) \quad | c_1< ph,q,n |R > + c_2< ph,q,n |T > |^2$$

for n =1,2, … which of course corresponds to effective interference patterns appearance in full agreement with well-known real experimental facts..

So, it is not hard to see, in full agreement with previous discussions, general concept of the effective collapse as spontaneous (non-dynamical) superposition or entanglement breaking (effective hiding) can be consistently applied in mentioned experiments (figure 4.) of the single photon interference at beam splitter. Namely, since wave packet approximations of all eigen states of PP “pointer” observable mutually extremely weakly interfere all conditions for effective appearance of the self-collapse at PP are satisfied. Simultaneously at the photon, via entanglement, relative collapse as effective quantum phenomena appears over eigen states of photon coordinate. It means that entangled state (32) turns out effectively (classically) spontaneously in statistical mixture of the states |ph,q,n >|PP,n,WP > of photon+PP with statistical weights (34) corresponding to interference patterns existence. Simultaneously, quantum exactly photon+PP is described by entangled quantum state (31).

Consider now experiment of single photon interference at beam splitter symbolically presented at figure 5. . It, in respect to experiment presented at figure 4., holds an additional quantum system, photon “trajectory entangler” TE, placed at “reflected” photon trajectory, between M1 and PP.

It can be supposed that TE can be such quantum system that satisfy the following conditions.. Suppose that TE hax three eigen states of some observable |TE,0 >, |TE,R > and |IET >. Suppose that TE is initially, i.e. before quantum dynamical interaction with photon, described by |TE,0 >. It implies that, immediately before quantum dynamical interaction between TE and photon, quantum super-system TE+photon is described by non-entangled quantum state |TR,0 >|ph >. Suppose that quantum dynamical interaction between TE and photon is described by unitary evolution operator **U(**TE+ph**)** defined by the following expressions

(35) **U(**TE+ph**)** |TE,0 >| R > = |TE,R > |R >

(36) **U(**TE+ph**)** |TE,0 >| T > = |TE,T > |T >

In this way photon “reflected” trajectory becomes correlated with |TE,R > and photon “transmitted” trajectory - with |TE,T >.Then, immediately after quantum dynamical interaction between TE and photon TE+photon turns out strictly deterministically from non-entangled quantum state |TE0 >|ph > in entangled quantum state

(37) |TE+ph > = **U(**TE+ph**)** |TE,0 >|ph > = $c_1$|TE,R > |R > + $c_2$ |TE,T >|T >

(We can simply demonstrate that such dynamical interaction without significant perturbation of the photon trajectories is possible in principle. Suppose that S emits single photon by quantum jump of some atom from excited state |E > in ground state

|G >. Suppose now that TE consists of two distant atoms of the same type as the atoms in S, where first (closer to M1) is initially in excited state while second (closer to PP) is initially in ground state. In this way TE is initially in quantum state |E > |G > . Obviously TE can immediately quantum dynamically interact only with "reflected" photon and this interaction can be simply described by stimulated emission of a new "copy" photon by first atom and absorption of this "copy" photon by second atom without any decoherence of the first photon "reflected" state. Simultaneously, after mentioned interaction, TE becomes described by quantum state |G > |E > . On the other hand, there is no any immediate quantum, dynamical interaction between TE and "transmitted" photon and non-interacting TD and "transmitted" photon stand in non-perturbed initial quantum states. It can be added that conceptually similar situation appears in real experiments of the interference of large molecules at diaphragm with two slits by trajectory detection by collisions wits small molecules [7]. It has been observed that here even by low intensive collisions, non-sufficient for significant perturbation of the large molecule trajectories but sufficient for realization of the entanglement between large and small molecule, there is effective absence of the interference patterns.)

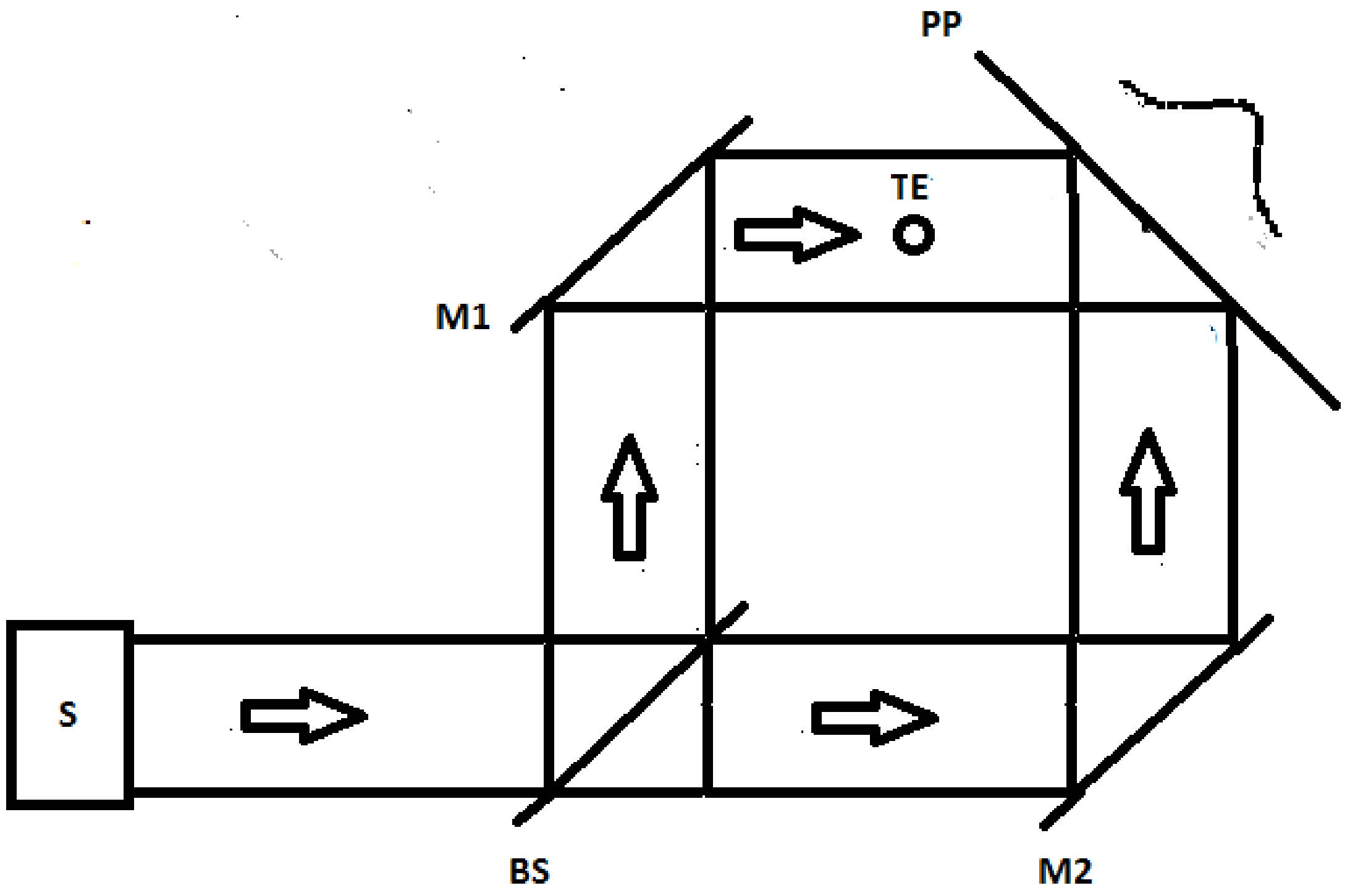

Figure 5.

So there is such quantum system(s) called photon trajetories "entangler" whose exact quantum dynamical interaction with photon yields exactly deterministically entanglement between quantum states of the "entangler" and corresponding photon trajectories. Simultaneously it means that mentioned exact quantum dynamical interaction does not yield exactly statistical mixture of mentioned quantum states of the "entangler" and photon. Simply spealing "entangler" is not "detector".

Suppose that after mentioned quantum dynamical interaction with photon, TE does not immediately quantum dynamically interact with photon or PP at all

Suppose however that PP before quantum dynamical interaction with TE+photon is described by quantum state (27) while simultaneously TE+photon (after quantum dynamical interaction between TE and photon) is described by entangled state (37). It implies that TE+photon+PP before quantum dynamical interaction between TE+photon and PP is described by partially entangled quantum state |TD+ph >|PP,0 > of TE+photon+PP

As it is not hard to see (according to suppositions that TE later does not immediately quantum dynamically interact with photon and PP) quantum dynamical interaction between TE+photon and PP can be presented by the following unitary evolution operator **1** x **U (**for **U** that satisfies (30) ). Then initial (before mentioned interaction) partially entangled quantum state |TE+ph >|PP,0 > quantum dynamically deterministically evolve in the following final (after mentioned interaction) quantum state of TE+photon+PP

(35) $(\mathbf{1} \times \mathbf{U}) |TE + ph > |PP,0 > =$
$= c_1|TE,R > \mathbf{U} |R > |PP,0 > + c_2 |TE,T > \mathbf{U} |T > |PP,0 > =$
$= \sum_{n,} (c_1|IE,R > < ph,q,n |R > + c_2 |TE,T >|< ph,q,n |T > ) |ph,q,n > |PP,n >$

Quantum state (35) can be further transformed in the following, correctly three-orthogonally expanded, entangled quantum state

(36) $\sum_{n,} a(n) |TDa(n) >|ph,q,n > |PP,n >$

where

(36) $a(n) = (|c_1|^2| < ph,q,n |R > |^2 + | c_2 |^2| <ph,q,n |T > |^2)^{1/2}$

represents a superposition coefficient and where

(37) $|TEa(n) > =.$
$(c_1|TE,R > < ph,q,n |R > + c_2|TE,T > < ph,q,n|ph,T > )/ a(n)$

represents a normalized quantum state of TE for n=1,2, … ..

Entangled quantum state ( 37) can be further partially wave packet approximated by entangled quantum state

(38) $\sum_{n,} a(n) |TDa(n) >|ph,q,n > |PP,n,WP >$

All this, according to previous discussions, admits that immediately after mentioned qantum dynamical interaction between TE+photon and PP at P**P** self-collapse spontaneously and effectively appears and correspondingly at TE+photon relative collapse (as effective quantum phenomena) appears. It other words entangled quantum state (38) effectively collapses or turns out spontaneously in statistical mixture of quantum states |TEa(n) >|ph,q,n > |PP,n,WP > with statistical weights

(39) $|a (n) |^2 = |c_1|^2| < ph,q,n |R > |^2 + | c_2 |^2| <ph,q,n |T > |^2$

for n=1,2, … .It also means that photon that photon is described effectively by effective statistical mixture of quantum st6ates ph,q,n > with identical statistical weights (39) for n=1,2, … . It of course excellently corresponds to absence of the interference paerns or to well-known real experimental facts. Here it is extremely important to be pointed out that statistical distribution (39),as it has veen previously discussed, is not caused by absolute collapse realized by previous interaction between TE and photon. It is result of the collapse as the spontaneous unitary symmetry, i.e. superposition breaking without breaking of the unit norm of quantum states.

All this simply implies that there is no any need for absolute collapse. As it is not hard to see, in case when "entangler" quantum states satisfy mentioned additional approximation conditions for spontanous superposition breaking "entangler" turns out in "detector". But even in this case"entangler"-"detector" realizes initially the collapse only effectively and this detection is quite compatible with PP final detection. In mentioned experiment, resuulta of PP final detection are effectively independent in case when "entangler"-"detector" is presented exactly as "entangler" ("actor") or effectively as "detector" ("spectzator"). It implies that here, i.e. 9n experiment presented symbolically at figure 5. as well as in its relized analogies [7], Schrödinger cat effect is realized. However, in all mentioned situations TE+photon+PP is finally exactly quantum mechanically described exclusively by entangled quantum state (36) ("however it unitary evolve", we can paraphrase Galileo Galilei).

Finally the following can be concluded. All exact quantum phenomena can be exactly and completely described by strictly deterministic unitary symmetric (that conserves superposition) quantum dynamical evolution (without any super-luminal phenomena) of quantum states (of the unit norm) in Hilbert space as basic physical space. In this way quantum mechanics is consistently founded. Detection or measurement can be consistently considered as a quantum-classical phase transition with spontaneous (non-dynamical) unitary symmetry (superposition) breaking (effective hiding).